\documentstyle[12pt,epsf]{article}

\catcode`\@=11
\long\def\@makefntext#1{
\protect\noindent \hbox to 3.2pt {\hskip-.9pt
$^{{\ninerm\@thefnmark}}$\hfil}#1\hfill}		

\def\@makefnmark{\hbox to 0pt{$^{\@thefnmark}$\hss}}  

\def\ps@myheadings{\let\@mkboth\@gobbletwo
\def\@oddhead{\hbox{}
\rightmark\hfil\ninerm\thepage}
\def\@oddfoot{}\def\@evenhead{\ninerm\thepage\hfil
\leftmark\hbox{}}\def\@evenfoot{}
\def\sectionmark##1{}\def\subsectionmark##1{}}

\renewcommand{\thefootnote}{\fnsymbol{footnote}}

\newcounter{sectionc}\newcounter{subsectionc}\newcounter{subsubsectionc}
\renewcommand{\section}[1] {\vspace*{0.6cm}\addtocounter{sectionc}{1}
\setcounter{subsectionc}{0}\setcounter{subsubsectionc}{0}\noindent
	{\normalsize\bf\thesectionc. #1}\par\vspace*{0.4cm}}
\renewcommand{\subsection}[1] {\vspace*{0.6cm}\addtocounter{subsectionc}{1}
	\setcounter{subsubsectionc}{0}\noindent
	{\normalsize\it\thesectionc.\thesubsectionc. #1}\par\vspace*{0.4cm}}
\renewcommand{\subsubsection}[1]
{\vspace*{0.6cm}\addtocounter{subsubsectionc}{1}
	\noindent {\normalsize\rm\thesectionc.\thesubsectionc.\thesubsubsectionc.
	#1}\par\vspace*{0.4cm}}

\newcounter{appendixc}
\newcounter{subappendixc}[appendixc]
\newcounter{subsubappendixc}[subappendixc]

\renewcommand{\appendix}[1] {\vspace*{0.6cm}
        \refstepcounter{appendixc}
        \setcounter{figure}{0}
        \setcounter{table}{0}
        \setcounter{equation}{0}
        \renewcommand{\thefigure}{\Alph{appendixc}.\arabic{figure}}
        \renewcommand{\thetable}{\Alph{appendixc}.\arabic{table}}
        \renewcommand{\theappendixc}{\Alph{appendixc}}
        \renewcommand{\theequation}{\Alph{appendixc}.\arabic{equation}}
        \noindent{\bf Appendix \theappendixc #1}\par\vspace*{0.4cm}}

\def\abstracts#1{{
	\centering{\begin{minipage}{12.2truecm}\footnotesize\baselineskip=12pt\noindent
	\centerline{\footnotesize ABSTRACT}\vspace*{0.3cm}
	\parindent=0pt #1
	\end{minipage}}\par}}

\renewenvironment{thebibliography}[1]
	{\begin{list}{\arabic{enumi}.}
	{\usecounter{enumi}\setlength{\parsep}{0pt}
\setlength{\leftmargin 1.25cm}{\rightmargin 0pt}
	 \setlength{\itemsep}{0pt} \settowidth
	{\labelwidth}{#1.}\sloppy}}{\end{list}}

\newcounter{itemlistc}
\newcounter{romanlistc}
\newcounter{alphlistc}
\newcounter{arabiclistc}

\newcommand{\fcaption}[1]{
        \refstepcounter{figure}
        \setbox\@tempboxa = \hbox{\footnotesize Fig.~\thefigure. #1}
        \ifdim \wd\@tempboxa > 6in
           {\begin{center}
        \parbox{6in}{\footnotesize\baselineskip=12pt Fig.~\thefigure. #1}
            \end{center}}
        \else
             {\begin{center}
             {\footnotesize Fig.~\thefigure. #1}
              \end{center}}
        \fi}

\newcommand{\tcaption}[1]{
        \refstepcounter{table}
        \setbox\@tempboxa = \hbox{\footnotesize Table~\thetable. #1}
        \ifdim \wd\@tempboxa > 6in
           {\begin{center}
        \parbox{6in}{\footnotesize\baselineskip=12pt Table~\thetable. #1}
            \end{center}}
        \else
             {\begin{center}
             {\footnotesize Table~\thetable. #1}
              \end{center}}
        \fi}

\def\@citex[#1]#2{\if@filesw\immediate\write\@auxout
	{\string\citation{#2}}\fi
\def\@citea{}\@cite{\@for\@citeb:=#2\do
	{\@citea\def\@citea{,}\@ifundefined
	{b@\@citeb}{{\bf ?}\@warning
	{Citation `\@citeb' on page \thepage \space undefined}}
	{\csname b@\@citeb\endcsname}}}{#1}}

\newif\if@cghi
\def\cite{\@cghitrue\@ifnextchar [{\@tempswatrue
	\@citex}{\@tempswafalse\@citex[]}}
\def\citelow{\@cghifalse\@ifnextchar [{\@tempswatrue
	\@citex}{\@tempswafalse\@citex[]}}
\def\@cite#1#2{{$\null^{#1}$\if@tempswa\typeout
	{IJCGA warning: optional citation argument
	ignored: `#2'} \fi}}

 1
 1
 1

\font\ninerm=cmr9

\begin{document}

\centerline{\normalsize\bf RELATIONS BETWEEN MONOPOLES, INSTANTONS }
\baselineskip=22pt
\centerline{\normalsize\bf AND  CHIRAL CONDENSATE \footnote{
Supported in part by FWF under Contract No.~P11456-PHY}}

\centerline{\footnotesize S.~THURNER, M.~FEURSTEIN, H.~MARKUM }
\baselineskip=13pt
\centerline{\footnotesize\it Institut f\"ur Kernphysik, TU Wien,
A-1040 Vienna, Austria}

\vspace*{0.9cm}

\abstracts{
We analyze the interplay of topological objects in four dimensional QCD.
The distributions of color magnetic monopoles obtained in the maximum
abelian gauge are computed around instantons in both pure and full QCD.
We find an enhanced probability of
encountering monopoles inside the core of an instanton on gauge field average. 
For specific gauge field configurations  we visualize the situation 
graphically.
Moreover we investigate how monopole loops and instantons are
locally correlated with the chiral condensate. 
}
\normalsize\baselineskip=15pt
\setcounter{footnote}{0}
\renewcommand{\thefootnote}{\alph{footnote}}

\section{Introduction and Theory}
There are two different kinds
of topological objects which seem to be important for the confinement mechanism:
color magnetic monopoles and instantons.
Color magnetic monopoles play the main role in the dual
superconductor hypothesis\cite{thooft1} where confinement occurs by
condensation of abelian monopoles via the dual Meissner effect.
There is strong evidence from lattice calculations that the idea of
dual superconductivity is in essence correct. 
On the other hand the role of instantons with respect to confinement is
not so clear. It is assumed that instantons can only cause confinement
in QCD if they form a so-called instanton liquid.\cite{SHU88}
In our lattice calculations we demonstrate that color magnetic monopoles 
and instantons are highly correlated.\cite{wir}
This might explain that both pictures of the 
confinement mechanism have the same  topological 
origin  and that both approaches can be united.

It is believed that both instantons and monopoles can explain chiral 
symmetry breaking.\cite{SHU88,MIA95} 
In this contribution we present first results on the local correlation 
functions  
of the chiral condensate, the topological charge density, and the monopole 
density. 

In order to investigate monopole currents one has to project $SU(N)$
onto its  abelian degrees of freedom, such that an abelian $U(1)^{N-1} $
theory remains.\cite{thooft2} This aim can be achieved by
various gauge fixing procedures. We employ the
so-called maximum abelian gauge which is most favorable for our
purposes.
For the definition of the monopole currents $m(x,\mu)$ we use the
standard method.\cite{SCH87} 
To extract abelian parallel transporters $u(x,\mu)$ 
after imposing the maximum abelian gauge one has to perform
the decomposition 
\begin{equation}
\tilde{U}(x,\mu) = c(x,\mu) u(x,\mu)  \ , \quad \mbox{\rm with} \quad (N=3)
\end{equation}
\begin{eqnarray} \label{ab_u}
u(x,\mu) &=& \mbox{\rm diag } [ u_{1}(x,\mu), u_{2}(x,\mu), u_{3}(x,\mu) ]
\ , \nonumber \\
u_{i}(x,\mu) &=& \exp \Big [ i \ \mbox{\rm arg } \tilde{U}_{ii}(x,\mu) -
\frac{1}{3} i \phi(x,\mu) \Big ] \ , \nonumber \\
\phi(x,\mu) &=& \sum_{i} \mbox{\rm arg } \tilde{U}_{ii}(x,\mu) \Big |
_{\mbox{\tiny mod $2\pi$}} \! \in \! (-\pi,\pi] . \nonumber
\end{eqnarray}
Since the maximum abelian subgroup $U(1)^{N-1}$ is compact, there
exist topological excitations. These are color magnetic monopoles which have
integer-valued magnetic currents on the links of the dual lattice:
\begin{equation}
m_{i}(x,\mu) = \frac{1}{2\pi} \sum_{\Box \ni \partial f(x+\hat\mu,\mu)}
\mbox{\rm arg } u_{i}(\Box) \ ,
\end{equation}
where $u_{i}(\Box)$ denotes a product of abelian links $u_{i}(x,\mu)$ around
a plaquette $\Box$ and $f(x+\hat\mu,\mu)$ is an elementary cube perpendicular
to the $\mu$ direction with origin $x+\hat\mu$.
The magnetic currents form closed loops on the dual lattice as a consequence
of monopole current conservation.
From the monopole currents we define the local monopole density as
$ \rho(x) = \frac{1}{ 3 \cdot 4V_{4}} \sum_{\mu,i} | m_{i}(x,\mu) | \ .  $
%

For the implementation of the topological charge on the lattice
there exists no unique discretization. In this work we
restrict ourselves to the so-called field theoretic definitions which
approximate the topological charge in the continuum, 
$
q(x)=\frac{g^{2}}{32\pi^{2}} \epsilon^{\mu\nu\rho\sigma}
\ \mbox{\rm Tr} \ \Big ( F_{\mu\nu}(x) F_{\rho\sigma}(x) \Big ) \ ,
$
in the following ways:\cite{divecchia}
\begin{equation}
 q^{(P,H)}(x)=-\frac{1}{2^{4}32\pi^{2}}
\sum_{\mu,\ldots =\pm 1}^{\pm 4}
\tilde{\epsilon}_{\mu\nu\rho\sigma} \mbox{\rm Tr} \ O_{\mu\nu\rho\sigma}^{(P,H)}
,
\end{equation}
with
\begin{equation}
O_{\mu\nu\rho\sigma}^{(P)} = U_{\mu\nu}(x) U_{\rho\sigma}(x) \ ,
\end{equation}
for the plaquette prescription and
\begin{eqnarray}
O_{\mu\nu\rho\sigma}^{(H)} &=&
U(x,\mu) U(x\!+\!\hat\mu,\nu) U(x\!+\!\hat\mu\!+\!\hat\nu,\rho) 
U(x\!+\!\hat\mu\!+\!\hat\nu\!+\!\hat\rho,\sigma) 
\nonumber \\ & \times &
U^{\dagger}(x\!+\!\hat\nu\!+\!\hat\rho\!+\!\hat\sigma,\mu) 
U^{\dagger}(x\!+\!\hat\rho\!+\!\hat\sigma,\nu)
U^{\dagger}(x\!+\!\hat\sigma,\rho) U^{\dagger}(x,\sigma), 
\end{eqnarray}
for the hypercube prescription.
We mention here that  the  topological charges
employed are locally gauge invariant, whereas the monopole currents are not.
The lattice and continuum versions of the theory represent
different renormalized quantum field theories, which differ 
by finite, non-negligible renormalization factors. 
A simple procedure  to get rid of renormalization constants,
while preserving physical information contained in lattice configurations,
is the cooling method.
The cooling procedure systematically reduces quantum fluctuations, 
and suppresses
differences between the different definitions of the topological charge.
In our investigation we have employed the so-called 
``Cabbibo--Marinari method''.

To measure correlations between topological quantities
we calculate the functions
\begin{eqnarray} \label{correlations}
\langle q(0) q(d) \rangle \ , 
\langle \rho(0) \rho(d) \rangle \ , 
\langle \rho(0) q^{2}(d) \rangle \ ,
\langle q^{2}(0) \bar \psi \psi (d) \rangle \ ,
\langle \rho(0) \bar \psi \psi (d) \rangle ,
\end{eqnarray}
which are normalized  after subtracting 
the corresponding cluster values.
Since topological objects with opposite sign are equally distributed,
we correlate the monopole density and the local 
chiral condensate with the square of the topological charge density.

\section{Results and Discussion}

\begin{figure*}[thb]
\begin{center}
\begin{tabular}{ccc}
\vspace{-0.8cm}
\\
\hspace{-0.8cm} \epsfxsize=5.3cm\epsffile{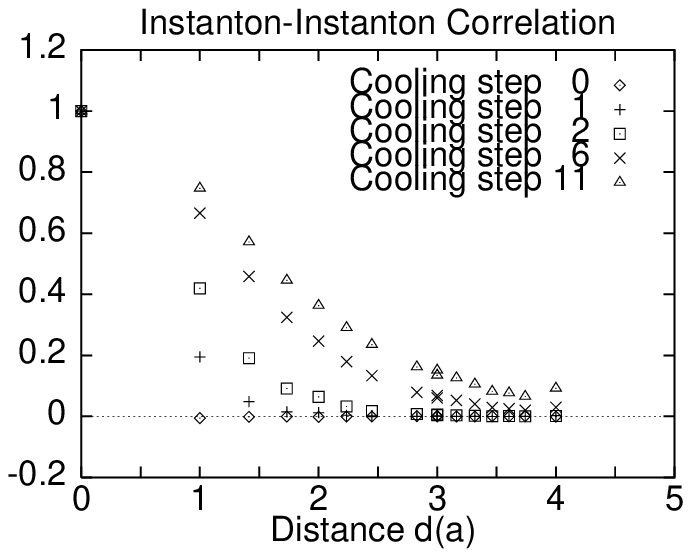}
\vspace{-1.6cm} & 
\hspace{-0.6cm} \epsfxsize=5.3cm\epsffile{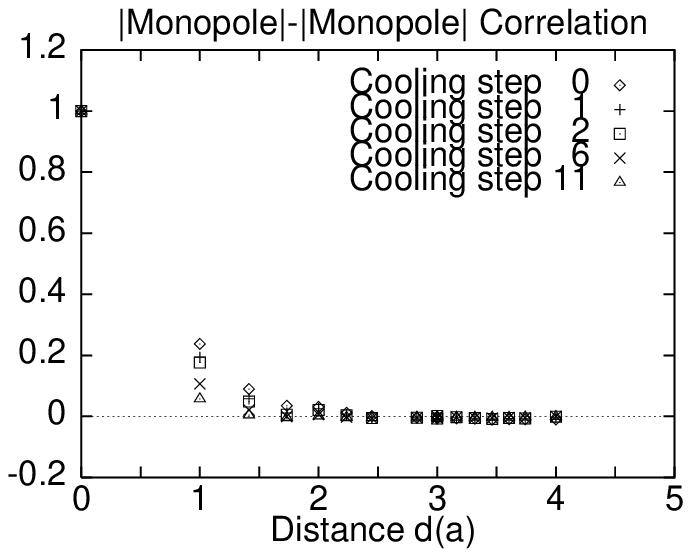}
&
\hspace{-0.6cm} \epsfxsize=5.3cm\epsffile{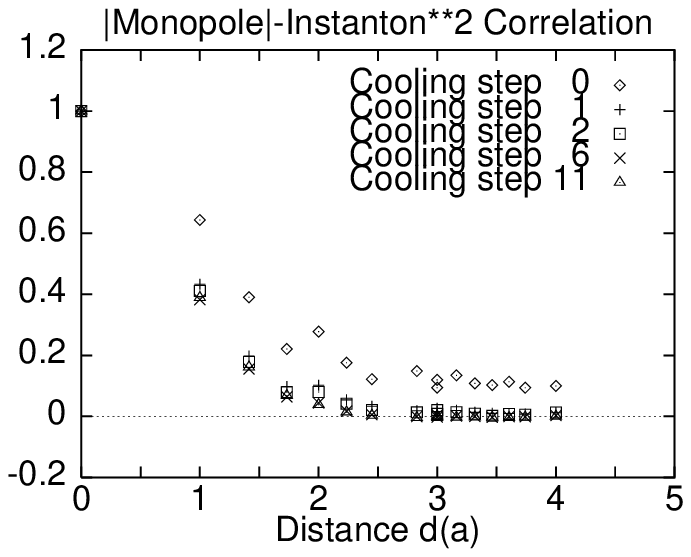} \\ 
{\footnotesize \hspace{3.10cm} (a)} & {\footnotesize \hspace{3.30cm} (b)}
 &{\footnotesize \hspace{3.30cm} (c)}\\
\vspace{0.7cm}
\end{tabular}
\end{center}
{\baselineskip=13pt
\small
Figure 1.~Correlation functions between topological charge densities
and monopole densities in the confinement phase ($\beta=5.6$) for
$0,1,2,6,11$ cooling steps.
The instanton autocorrelations (a) grow with cooling
reflecting the existence of extended instantons whereas the monopole
autocorrelations (b) decrease since monopoles become diluted.
The correlations between monopoles and instantons (c) are hardly affected by 
 cooling with a range of approximately
two lattice spacings indicating a nontrivial relation between
these topological objects.
\baselineskip=15pt}
\end{figure*}

Our simulations for the $SU(3)$ case 
were performed on an $8^{3} \times 4$ lattice with
periodic boundary conditions using the Metropolis algorithm.
For pure QCD we evaluated the path integral 
with standard Wilson action in the confinement
phase at inverse gluon coupling $\beta = 6/g^2= 5.6$. 
The  measurements were taken
on 1000 configurations separated by 50 sweeps.
Each configuration was cooled and then subjected to 300 gauge fixing
steps enforcing the maximum abelian gauge.
Switching on dynamical fermions we simulated full QCD with 3 flavors
of Kogut-Susskind quarks of equal mass $ma=0.1$. We performed  runs
in the  confinement at $\beta=5.2$ 
and measured on 1000 configurations separated by 50 sweeps.

The correlation functions between topological quantities in pure QCD 
(hypercube definition for $q$)
are shown in Fig.~1 for several cooling steps.
The range of the 
instanton autocorrelation $qq$ (a)
being originally
$\delta$-peaked grows rapidly with cooling reflecting the occurance of
extended instantons.
In contrast the $\rho \rho$-correlation (b) decreases
since monopole loops become dilute with cooling. The
$\rho  q^{2}$-correlation (c) seems rather insensitive to cooling and clearly
extends over more than two lattice spacings, indicating some nontrivial
local correlation between monopoles and topological charges.

Fig. 2 shows correlation functions of full QCD 
in the confinement region.  
The  $\rho q^2$ correlation (a) looks similar to the corresponding function in 
pure QCD (Fig. 1a). The same holds for the instanton and monopole 
autocorrelation functions (not shown). 
In the case of the $\bar \psi \psi q^2$ correlation (b) 
exponential fits show that an increasing number of cooling steps 
results in a narrower  correlation function. 
The $\bar \psi \psi \rho$ correlation (c) on the other hand is not sensitive 
to cooling  and has the same exponential decay as the 
$\bar \psi \psi q^2$ correlation after some cooling steps. 

\begin{figure*}[thb]
\begin{center}
\begin{tabular}{ccc}
\vspace{-0.5cm}
\\
\hspace{-1cm} \epsfxsize=5.6cm\epsffile{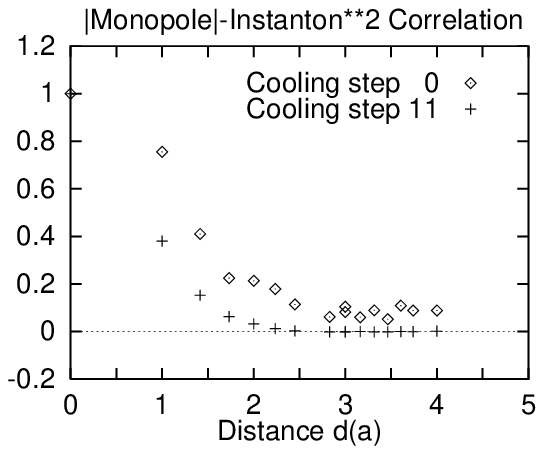} 
\vspace{-1.8cm} 
&
\hspace{-0.80cm} \epsfxsize=5.6cm\epsffile{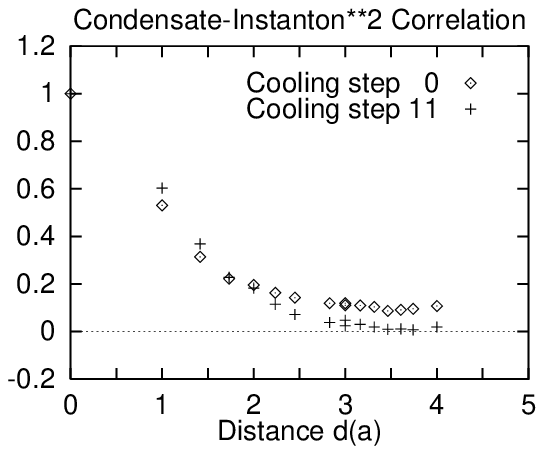}
&
\hspace{-0.90cm} \epsfxsize=5.6cm\epsffile{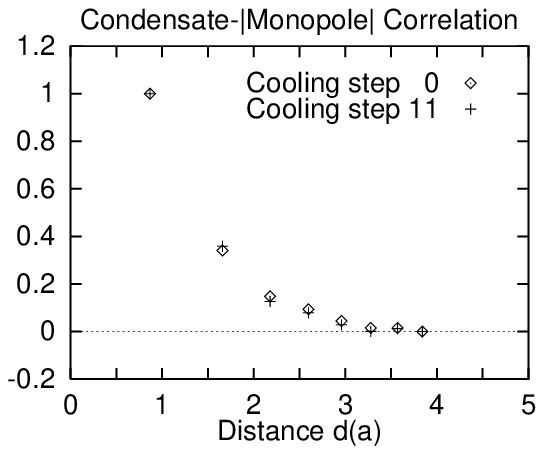}
\\ 
{\footnotesize \hspace{2.80cm} (a)} & {\footnotesize \hspace{3.00cm} (b)}
 &{\footnotesize \hspace{3.00cm} (c)}\\
\vspace{1.0cm}
\end{tabular}
\end{center}
{\baselineskip=13pt
\small
Figure 2.~ Correlation functions in the presence of dynamical 
           quarks in the confinement ($\beta=5.2$). The monopole-instanton 
           correlation (a) is almost the same as in pure $SU(3)$. The 
           correlation of the quark condensate and the topological charge (b) 
           is cooling dependent, whereas the correlation between the condensate 
           and the monopole density is not. All correlations extend 
           over two lattice spacings and indicate local 
           correlations of the chiral condensate and topological objects.
\baselineskip=15pt}
\end{figure*}

To obtain some insight into the topological correlations, 
Fig. 3 presents a cooling history of an $SU(2)$  
gluon field at a fixed  time slice on a $12^3 \times 4$ lattice.
The topological charge density using the plaquette and the
hypercube definition is displayed for cooling steps  0, 15 and 25.
A dot is plotted if $|q(x)| >0.01$. 
The lines represent the monopole loops.
Without cooling the topological charge distribution cannot be resolved 
from the  noise.
Also the monopole loops do not exhibit a  structure.
After 15-20 cooling steps one can identify clusters of topological
charge with instantons. 
At cooling  steps 35-40 the instanton and antiinstanton 
begin to approach each
other until they  annihilate several cooling steps after (not shown).
Monopole loops also thin out with cooling, but they survive in the
presence of instantons. In general, there is
 an enhanced probability that monopole loops exist  in the vicinity of
instantons.
\begin{figure*}[tb]
\begin{center}
\begin{tabular}{ccc}
\\
{\small Cooling step 0} & {\small Cooling step 15} &{\small Cooling step 25} \\
\hspace{-0.4cm} \epsfxsize=5.0cm\epsffile{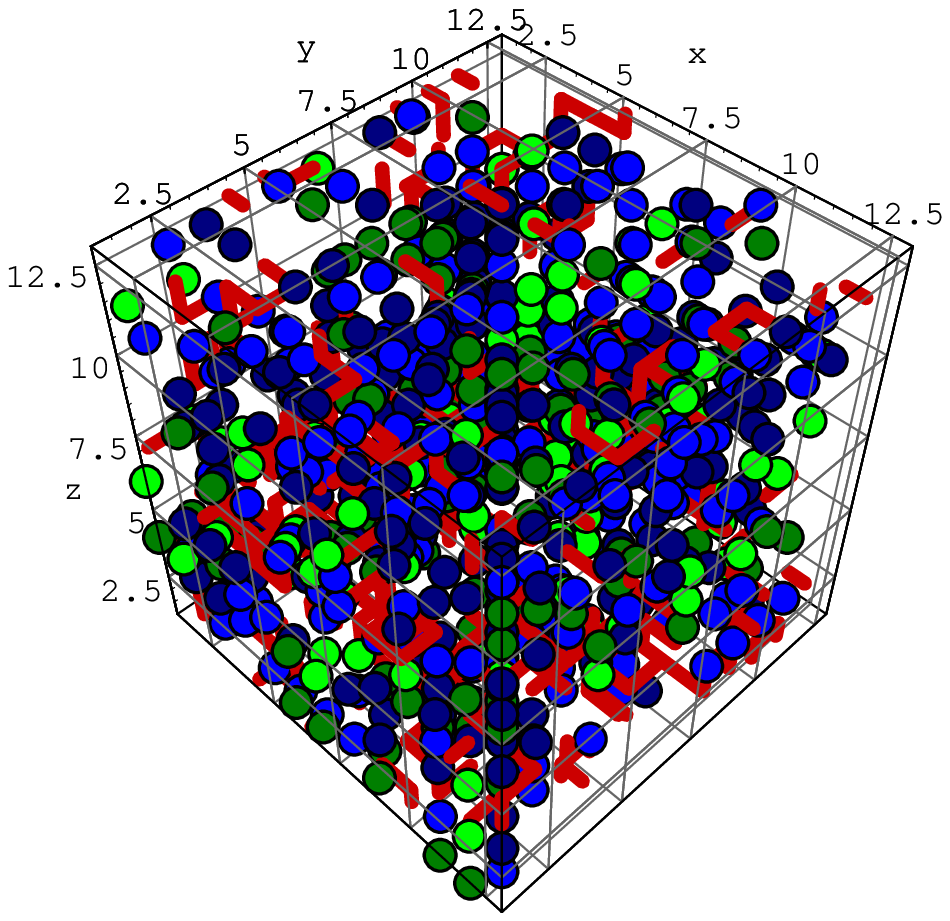}
&
\hspace{-0.3cm} \epsfxsize=5.0cm\epsffile{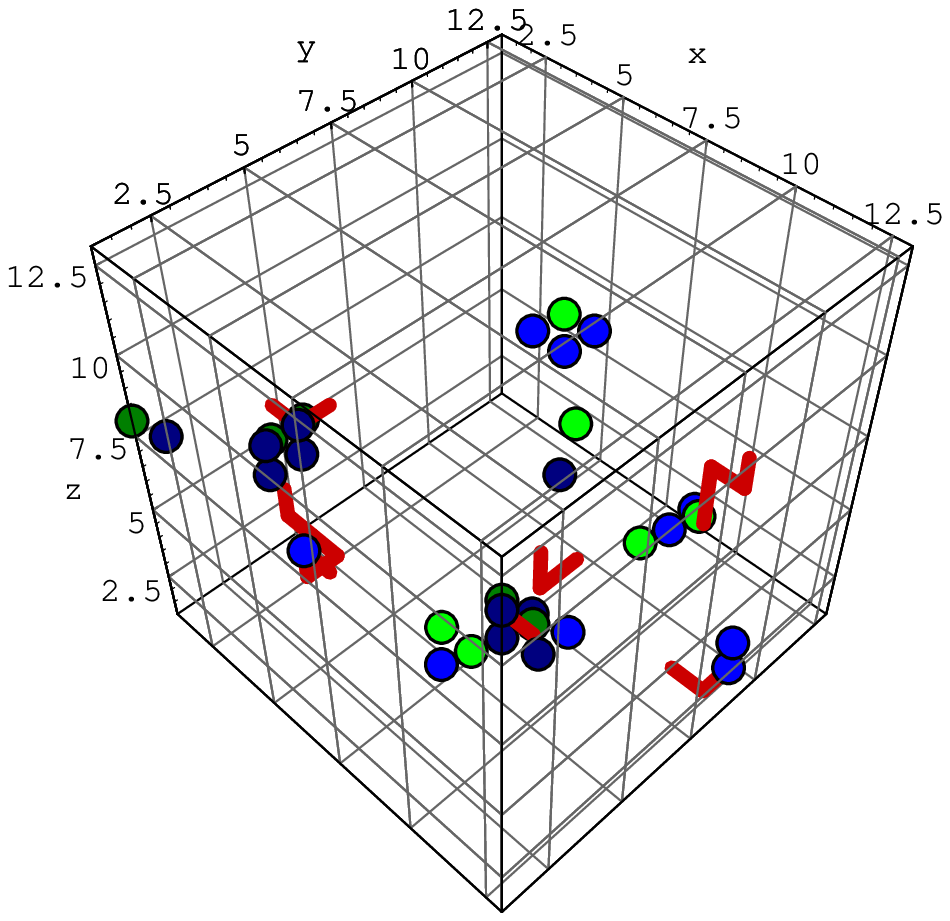}
&
\hspace{-0.3cm} \epsfxsize=5.0cm\epsffile{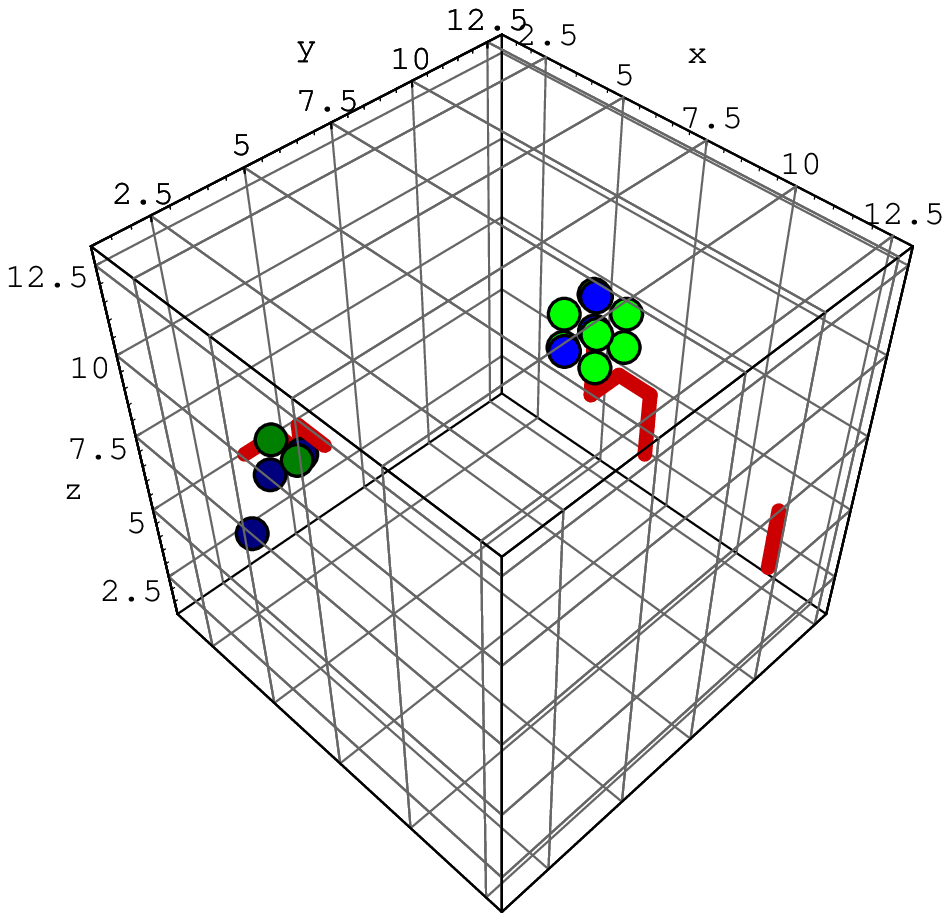}
\\
\end{tabular}
\end{center}
{\baselineskip=13pt
\small Figure 3.~Cooling history  for a time slice of a single
gauge field
configuration. The dots represent the topological  charge distribution. 
Monopole loops are represented by  lines. It can be seen that with cooling
instantons evolve from noise  accompanied by monopole loops
in almost all cases. Note that the black-and-white pictures do not present 
the situation so clearly as color plots. 
\baselineskip=15pt}
\end{figure*}

In summary, our calculations of correlation functions 
between topological objects 
and the chiral condensate yield a 
range of about two lattice spacings. This  suggests  that  the 
chiral condensate takes a nonvanishing value predominantly in the regions  of 
instantons and monopole loops. To our knowledge this observation is the 
first direct indication that chiral symmetry breaking occurs 
locally in the vicinity  of nontrivial topological structure. Our calculations 
are in agreement with other studies in $SU(2)$.\cite{MIA95} 
With the visualization of instantons and monopole loops in specific gauge field 
configurations we show directly that at the sites of instantons monopole loops 
are present. 
This confirms our conjecture  that monopoles and 
instantons might be two faces of a more subtle fundamental topological 
object, which even might carry an electric charge.\cite{dyons96}   


\end{document}